\documentclass[12pt]{iopart}

\usepackage{graphicx}
\usepackage[latin1]{inputenc}
\usepackage{amsmath}
\usepackage{xcolor}
\usepackage[urlcolor=blue]{hyperref}      % links to citations
\hypersetup{
    colorlinks = true,                    % text and not border
    citecolor = {blue},
    linkcolor = {purple},
           }
\begin{document}

\title{A devil's advocate view on `self-organized' brain criticality}
\author{Claudius Gros}

\address{Institute for Theoretical Physics, Goethe University Frankfurt, Germany}
\ead{gro07[@]itp.uni-frankfurt.de}
\vspace{10pt}
\begin{indented}
\item[]January 2021
\end{indented}

\begin{abstract}
Stationarity of the constituents of the body and of 
its functionalities is a basic requirement for life, 
being equivalent to survival in first place. Assuming
that the resting state activity of the brain serves 
essential functionalities, stationarity entails that 
the dynamics of the brain needs to be regulated on a 
time-averaged basis. The combination of recurrent 
and driving external inputs must therefore lead to a
non-trivial stationary neural activity, a condition 
which is fulfilled for afferent signals of varying 
strengths only close to criticality. In this view,
the benefits of working vicinity of a second-order
phase transition, such as signal enhancements, are 
not the underlying evolutionary drivers, but side 
effects of the requirement to keep the brain functional 
in first place. It is hence more appropriate to use the 
term `self-regulated' in this context, 
instead of `self-organized'.
\end{abstract}

%\maketitle

%%%%%%%%%%%%%%%%%%%%%%%%%%%%%%%%%%%%%%%%%%%%%%%%%%%%%%%%%
%%%%%%%%%%%%%%%%%%%%%%%%%%%%%%%%%%%%%%%%%%%%%%%%%%%%%%%%%

%===========================
\section{Life as a stationary flow equilibrium}
%===========================

A plethora of definitions may be used to characterize 
what it means to `live' \cite{luisi1998various}. 
Putting aside philosophical niceties, it is clear 
that survival is guaranteed on a day-to-day basis 
only when a certain constancy of the body and its 
core functionalities is achieved. Here we will lay 
out a compact line of arguments on how this type 
of `constancy' can be quantified in terms of a
stationary flow equilibrium. In particular we will 
argue that statistical stationarity is a condition sine 
qua non for the brain and that it necessarily implies 
a modus operandi close to criticality. Critical, or 
near-critical brain dynamics is in this perspective 
nothing more than a particular aspect of the demand
to retain stationary functionalities, here the 
resting state activity. This does not rule out 
secondary advantages of operating close to second-order
phase transition, like improved information processing 
\cite{bertschinger2004real,schubert2020local}. 

Our considerations are embedded in the ongoing discussion 
of experimental evidences regarding 
critical \cite{beggs2003neuronal},
quasi-critical \cite{fosque2020evidence}, or
sub-critical \cite{priesemann2014spike} 
neuronal brain dynamics, which include also
alternative explanations for the observed non-universal 
power laws \cite{markovic2014power}. Overall,
the interaction between theory and experiments
is in a state of fluid progress \cite{munoz2018colloquium}.
We will start with a general definition of stationary 
processes, which is then applied and illustrated in a first 
step to two problems from the neurosciences, Hebbian learning 
and spectral radius regulation. The latter will lead in 
a subsequent step to the control of overall brain activity, 
and with it to brain criticality. 

%\cite{droste2013analytical}
%{Analytical investigation of self-organized criticality in neural networks},
% look at references

%\cite{munoz2018colloquium}
%{Colloquium: Criticality and dynamical scaling in living systems},
%[[it is important to compare to experiments, return times
%of random walks, area covered]]

%---------------------------
\subsection{Stationarity of essential functionalities}
%---------------------------

Everything is in motion, ``panta rhei'' as the 
Greek philosopher Heraclitus expressed it 
\cite{bowe2005heraclitus}. This in particular 
true for higher lifeforms, which are characterized 
by high metabolic turnover rates. Consider, f.i.,
the continued recycling of proteins occurring on 
a daily rhythm in our cells. Once synthesized, 
protein have a finite lifetime, which ranges from 
hours to weeks \cite{mathieson2018systematic},
with typical half lifes of about twenty hours 
\cite{boisvert2012quantitative}. Half of us is 
gone when we look into the mirror the next day.
But we do not notice, generically speaking, because 
our cells work in a stationary flow equilibrium, 
continuously regenerating degenerated proteins. 
Constancy of biological features is obtained hence
in large part by self-regulated flow equilibria.
A specific example are the spines of neuronal 
cells, which can persist for 2-3 weeks in the adult 
CA1 of the hippocampus 
\cite{attardo2015impermanence}, and even longer 
in the neocortex \cite{holtmaat2005transient}.
This notwithstanding that the constituent proteins, 
including synapse-forming membrane complexes 
\cite{cohen2019neuronal}, have substantially 
shorter life times.

%---------------------------
\subsection{Statistical stationarity}
%---------------------------

So far, we did discuss stationarity in general
terms. The statistical aspect can be captured
to first order by simple probability distributions,
like
\begin{equation}
p_t(s) = \frac{1}{T} \int_{t-T}^t dt' \delta(s-S(t'))\,,
\label{p_s}
\end{equation}
which measures the probability that the 
observable $S=S(t)$ takes the value $s$ 
in the interval $[t-T,t]$, with $T$
being the observation period. The 
observable could be, f.i.\ the
interspike interval, or the neural
activity, the later when dealing 
with rate encoding neurons. A system
is stationary, in a strict sense, when 
$p_t(s)$ is invariant with respect to 
time $t$. In practice, time invariance 
will be satisfied only approximately.
Another caveat are the diverging time
scales that appear when closing in to a
second order phase transition. At this
point the observation period should also
be taken to diverge, $T\to\infty$,
strictly speaking. The same caveat holds 
when attempting to tune a system slowly 
to a critical point \cite{tredicce2004critical}.

Of course, instead of the distribution
of a single scalar variable, one could 
consider the cross-correlation between
neural activities, or other non-trivial
or higher-order statistical ensembles. 
For the purpose of the present article,
$p_t(s)$ is sufficient. Next we illustrate 
how the concept of statistical stationarity
can be put to work.

%--------------------------------------
\begin{figure}[t]
\centerline{
\includegraphics[width=0.9\textwidth]{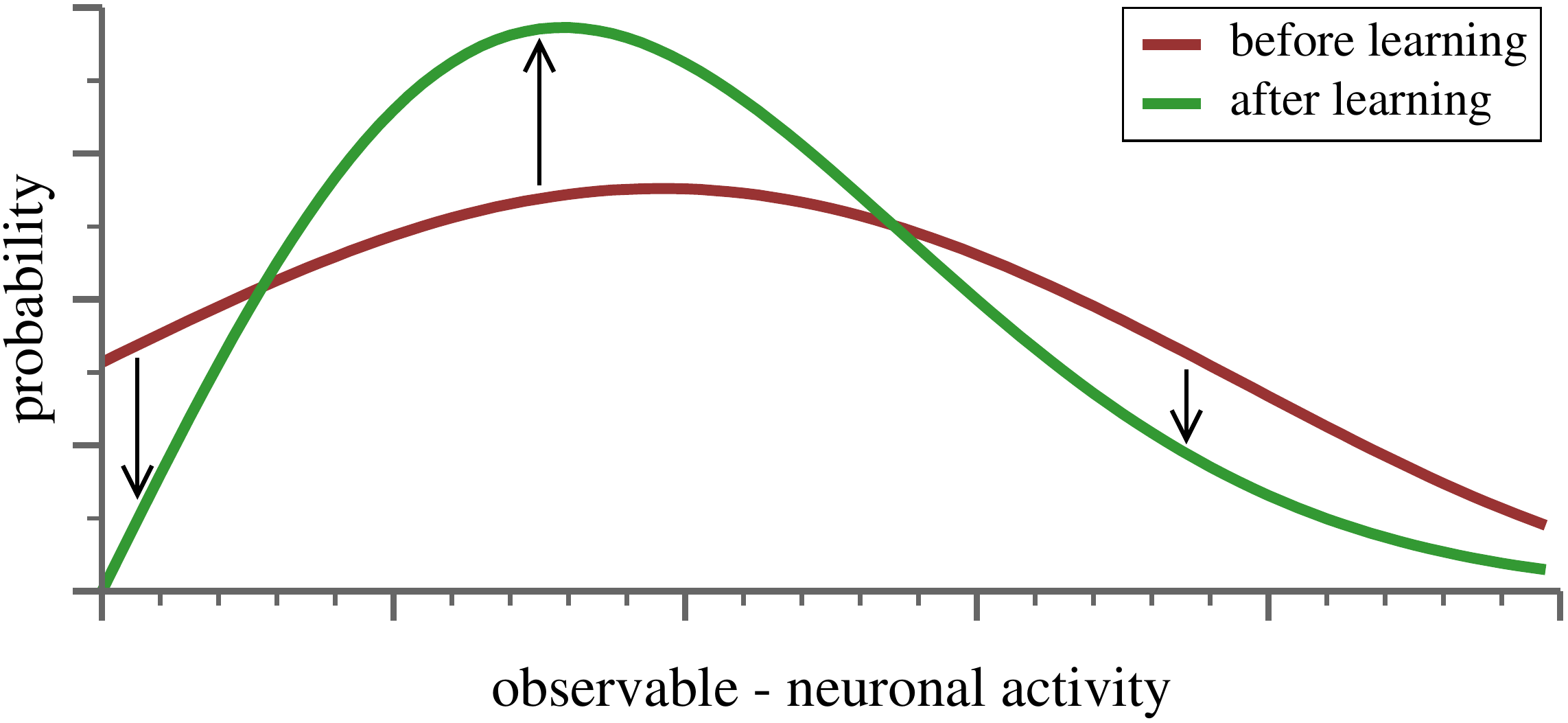}
           }
\caption{{\bf Learning induces stationarity.}
A neuron subject to inputs drawn from a given
ensemble of patterns produces firing patterns with 
a certain statistics (`before learning'), as
defined by (\ref{p_s}). During learning, the 
statistics of the firing pattern changes (arrows).
Once learning is completed, the neuronal activity
becomes stationary (`after learning'). Reversely, 
neuronal learning rules can be derived by optimizing 
a suitable objective function for stationarity,
such as the Fisher information (\ref{Fisher}).
}
\label{fig_HebbianLearning}
\end{figure}
%--------------------------------------

%===========================
\section{Hebbian learning as a `side effect' of stationary activity}
%===========================

The stationarity principle is not just an 
abstract, high-level concept. Instead, it
is relevant on a definitively practical level. 
As an example we show that the stationary
principle can be used to derive concrete 
expressions for a fundamental process,
Hebbian-type learning.

The framework is deceptively simple. Consider
a rate encoding neuron, for which the distribution
of outputs $s$ is given by $p_t(s)$, as defined
by (\ref{p_s}). The neuron learns, forming a 
receptive field by adapting its synaptic weights 
$w_i$ in response to the ensemble of input patterns 
received \cite{miller1994role}. As usual one assumes
that the statistics on the afferent activity patterns 
is stationary \cite{brito2016nonlinear}. 

During learning, the afferent synaptic weights $w_i=w_i(t)$ 
change, and with it the distribution $p_t(s)$
of the output activity. Synaptic weights will 
cease to change once the receptive field has 
formed, viz when learning is complete. At this 
point also $p_t(s)$ stops to change, becoming 
stationary, as illustrated in
Fig.~\ref{fig_HebbianLearning}. Stationarity is 
hence a consequence of Hebbian learning.

Reversely, given that stationarity results from 
Hebbian learning, one can derive Hebbian learning
rules by optimizing suitable information-theoretical
objective functions for stationarity 
\cite{echeveste2014generating}. A key candidate
is the Fisher information \cite{gros2015complex},
which takes the general form
\begin{equation}
F_\Theta = \int ds\, p_t(s)
\Big[\Theta\ln(p_t(s))\Big]^2,
\label{Fisher}
\end{equation}
where $\Theta$ is a differential operator
with respect to a quantity of interest.
As an example consider with
\begin{equation}
\Theta \,\to\, \frac{\partial}{\partial w_i}
\label{Theta_w_i}
\end{equation}
the derivative relative to a specific afferent
synaptic weight $w_i$. The Fisher information
(\ref{Fisher}) then measures how $p_t(s)$
changes when $w_i$ is modified, which is
exactly what happens during Hebbian learning.

Using (\ref{Fisher}) in conjunction with
(\ref{Theta_w_i}) as an objective function
for Hebbian learning is not suitable, with
the reason being that inter-synaptic competition,
the driver for receptive field formation, would
be absent. An alternative is the scalar
differential operator
\begin{equation}
\Theta \,\to\, \sum_i w_i\,\frac{\partial}{\partial w_i}\,,
\label{Theta_w_all}
\end{equation}
for which the respective Fisher information 
(\ref{Fisher}) incorporates the sensitivity
of $p_t(s)$ with respect to all synaptic weights
on an equal footing. Indeed, it can be 
shown that non-linear Hebbian learning rules
are obtained by minimizing $F_\Theta$, when
$\Theta$ is given by (\ref{Theta_w_all}).
The obtained learning rules do what all Hebbian
learning rules do, a principal component
analysis, being at the same time self-limiting
\cite{echeveste2014generating,echeveste2015fisher}.

The previous consideration show that activity regulation
is more than just another homeostatic process.
Given that one derive core synaptic plasticity
rules, it is evident that the stationarity of 
neural activity can be considered to play
the role of a first principle.
This does not rule out the significance
to optimize alternative objective functions, f.i.\
entropy for intrinsic plasticity \cite{triesch2007synergies},
mutual information within linear networks \cite{linsker1992local},
and for spike-time dependent plasticity \cite{chechik2003spike}.
It has been argued in this context, that
self-organizing systems may be `guided' 
with the help of a range of competing
generating functionals \cite{gros2014generating},
one for every degree of freedom.

%--------------------------------------
\begin{figure}[t]
\centerline{
\includegraphics[width=0.6\textwidth]{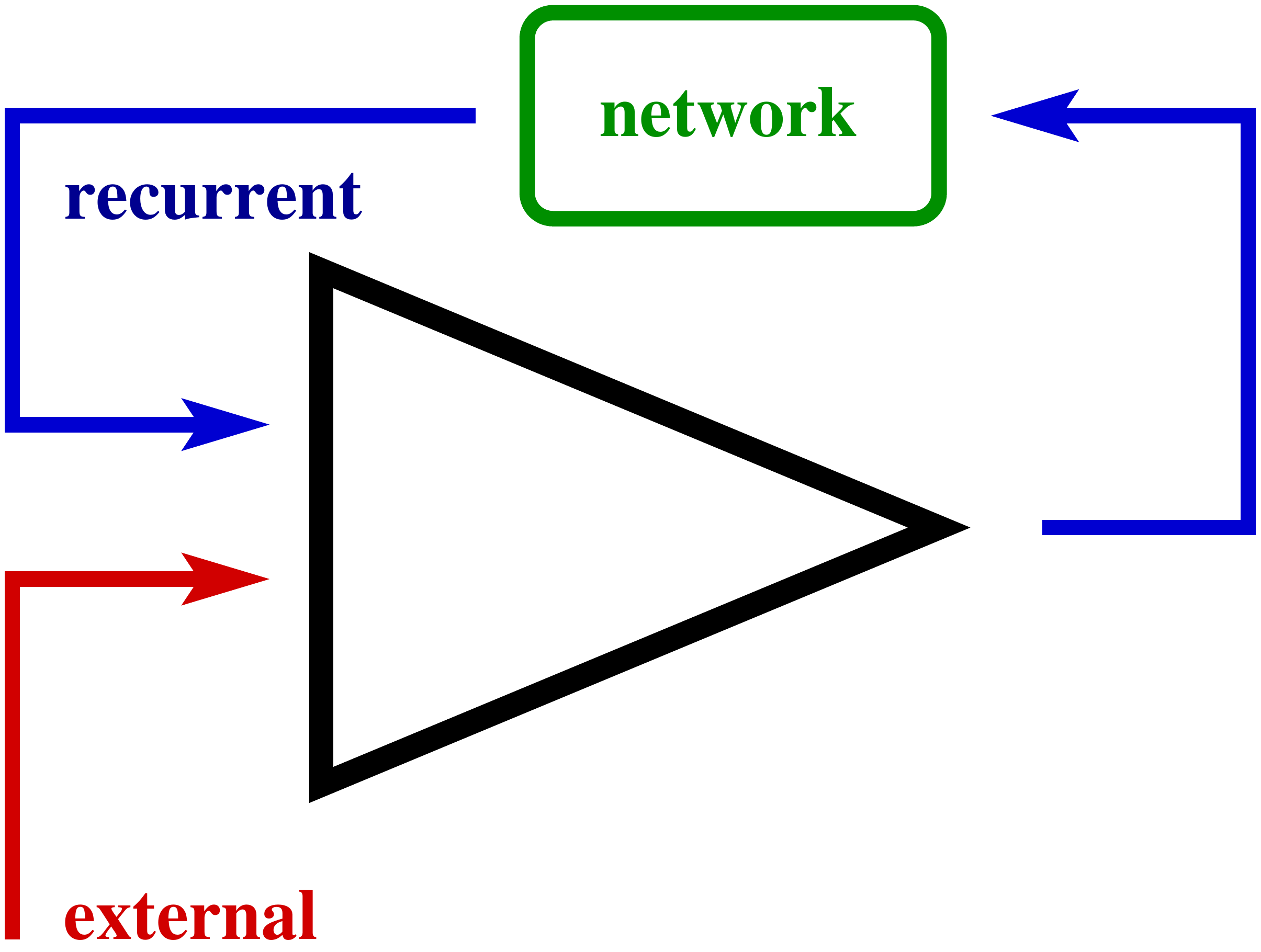}
           }
\caption{{\bf Flow of activity through a neuron.}
A neuron (black), receiving both external (red), and recurrent 
inputs (blue). Both inputs combined generate an
output, which in turn induces indirectly, via other
neurons (green), the recurrent input. The neuron is 
part of the overall network, with the spectral 
radius $R_w$ of the network determining the magnitude
of the recurrent input, in terms of the neuron's
output. Regulating the activity flow,
(recurrent input) $\to$ (output), in the presence
of external input, allows neurons to regulate
locally a global quantity, the spectral radius.
}
\label{fig_activityFlow}
\end{figure}
%--------------------------------------

%===========================
\section{Stationary activity flows\label{sect_activity_flows}}
%===========================

As a first step towards understanding the role
of the stationary condition for brain criticality
we pointed in the previous section to the
interplay between stationary activity and learning.
Now we turn to the interrelation between the
flow of activity through individual neurons and
the spectral radius of the recurrent synaptic weight
matrix. The latter determines in turn the closeness
to the critical point. 

The eigenvalues $\lambda_\alpha$ of a real, but 
non-symmetric matrix $\hat{W}$ are in general 
complex. The spectral radius $R_w$ of $\hat{W}$
\begin{equation}
R_w  = \max_{\alpha} |\lambda_\alpha|,
\qquad\quad
\hat{W}\mathbf{e}_\alpha = \lambda_\alpha\mathbf{e}_\alpha,,
\label{R_w}
\end{equation}
is given by the largest eigenvalue in absolute terms.
The matrix in question is here the synaptic weight 
matrix, $(\hat{W})_{ij} = w_{ij}$. For small activities, 
in the linear regime, the type of recurrent dynamics
generated by $\hat{W}$ is uniquely determined by
the spectral radius:
\begin{equation}
R_w \ :\  \left\{\begin{array}{rcl}
< 1 &:& \mbox{subcritical} \\
= 1 &:& \mbox{critical} \\
> 1 &:& \mbox{chaotic} 
\end{array}\right.\,.
\label{R_w_critical}
\end{equation}
Activity dies out in the subcritical regime
when the network is isolated, viz when there
is no driving external input. Neural activity
starts a runaway growth in contrast when 
$R_w>1$, which is however limited by the 
non-linearity of the transfer function. The 
resulting state is then chaotic, given that
$R_w$ determines whether the flow contracts
or expands, when time is discrete
\cite{gros2015complex,wernecke2019chaos}, as 
assumed here. For continuous time the system is instead 
critical when the maximal Lyapunov is zero
\cite{gros2015complex}.

Our considerations are based, as stated above, 
on models with discrete time, but this is not a 
restriction. The reason is that the overall 
flow of activity through a neuron 
is governed by a propagator, $P_{\Delta t}$,
which maps the input the neuron receives at a 
given time $t$ to the input received later, 
at time $t+\Delta t$. The exact sampling frequency 
$1/\Delta t$ is not relevant, as long as it reflects
the time scales of the involved biological processes,
which may involve refractory periods \cite{moosavi2017refractory}.
In general the spectral radius of $P_{\Delta t}$ 
is to be considered, and not of the bare synaptic weight 
matrix $\hat{W}$, as done above for simplicity. 
Both are however identical, in the linear regime, 
the situation discussed here, for which eventual 
neuronal gains act functionally as synpatic scaling 
factors.

%---------------------------
\subsection{Spectral radius regulation}
%---------------------------

In practice, neurons receive both recurrent
and external inputs, as illustrated in
Fig.~\ref{fig_activityFlow}. The output
of a given neuron flows through the
embedding network, becoming in the end
the recurrent input. On the average,
the activity generated by the neuron
considered is rescaled on its way 
through the surrounding neurons by 
the spectral $R_w$. This is because all
other components of the activity are 
subleading in an eigenspace decomposition
of the synaptic weight matrix, since 
$|\lambda_\alpha|\le R_w$. Stationarity 
is achieved when 
\begin{equation}
(\mbox{recurrent\ input})^2 \ \equiv \ 
R_w^2\,(\mbox{output})^2
\label{flow_stationary}
\end{equation}
holds as a suitable time averages. Given a
target spectral radius $R_w$, neurons can 
satisfy the flow condition (\ref{flow_stationary})
by regulating their gain \cite{schubert2020local},
which corresponds to a rescaling of the afferent
synaptic weights. This is quite remarkable, as it 
implies that a global quantity, the spectral radius, 
can be regulated by relying solely on local information. 

The adaption rule resulting from (\ref{flow_stationary}),
which has been termed `flow control' 
\cite{schubert2020local}, is effectively
based on the circular law of random matrix 
theory, which states that the spectral radius of a 
matrix with uncorrelated entries is given by
the variance of its elements \cite{tao2008random}.
The circular law states also that the eigenvalues of
a real but non-symmetric matrix $\hat{W}$ are 
uniformly distributed in the complex plain on a disk
with radius $R_w$. This implies that most eigenvalues
are smaller in magnitude. Note that flow control is an 
online rule, functioning while the system is operative, 
viz while processing a continuous stream of external 
inputs \cite{schubert2020local}.

%--------------------------------------
\begin{figure}[t]
\centerline{
\includegraphics[width=0.9\textwidth]{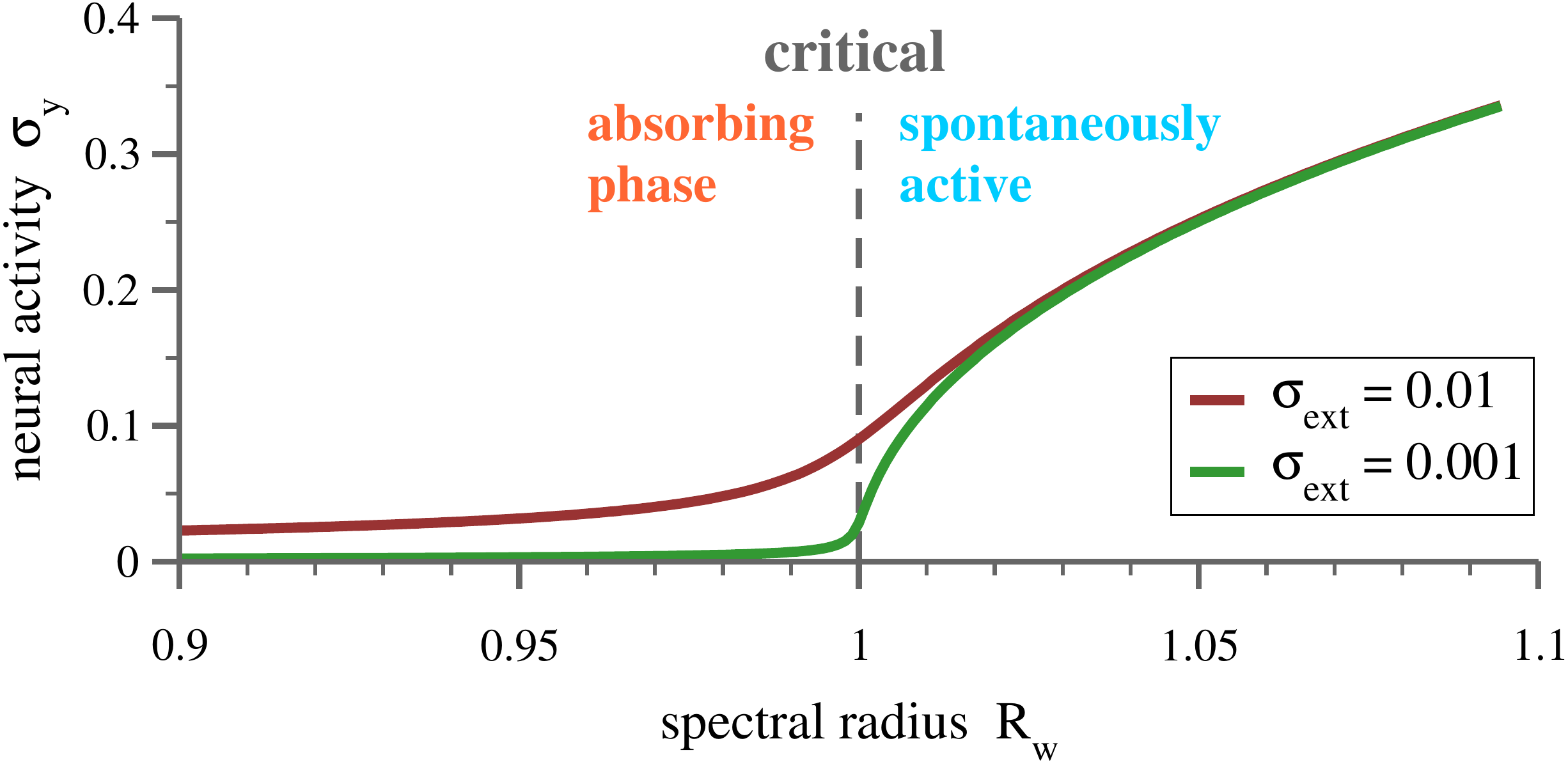}
           }
\caption{{\bf Brain criticality as an absorbing phase transition.}
The solution of the self-consistency condition (\ref{sigma_y}),
which describes the activity of a neural net with spectral 
radius $R_w$ in terms of the standard deviation $\sigma_y$
of the neuronal activity. 
Also present is an external driving having a standard 
deviation $\sigma_{\rm ext}$. Shown are the cases
$\sigma_{\rm ext}=10^{-2}/10^{-3}$. For vanishing input,
$\sigma_{\rm ext}\to0$, the classical transition from an absorbing
(inactive) phase to an autonomously active (chaotic) state 
is recovered.
}
\label{fig_APT_activity}
\end{figure}
%--------------------------------------

%---------------------------
\subsection{Input induced activity}
%---------------------------

A network of rate-encoding subject to external
inputs will settle into a continuously active 
dynamical state. The input is characterized 
typically by the variance $\sigma_{\rm ext}^2$,
which is taken here to subsume both presynaptic 
activities and the weights of the afferent 
synapses. The steady-state variance $\sigma_y^2$ 
of the neural activity is then determined by 
the self-consistency condition \cite{schubert2020local}
\begin{equation}
2 R_w^2 (1-\sigma_y^2)^2\sigma_y^2 =
1-(1-\sigma_y^2)^2(1+2\sigma_{\rm ext}^2)\,,
\label{sigma_y}
\end{equation}
where $R_w$ is spectral radius of the synaptic 
weight matrix. Compare Fig.~\ref{fig_activityFlow}.
The derivation of (\ref{sigma_y}) is based on the 
assumption that inter-site correlations can be 
neglected, together with the Gaussian approximation 
$\tanh^2(x)\approx 1-\exp(-x^2)$ for the squared 
neuronal transfer function, taken here to be $\tanh()$. 
This approximation to the transfer function is 
similar in spirit to the use of the error 
function as the neural transfer function 
\cite{echeveste2015fisher}. Expanding in small 
variances $\sigma_y^2$ and $\sigma_{\rm ext}^2$,
one finds
\begin{equation}
\sigma_y \ \sim \ \left\{\begin{array}{rcl}
\sigma_{\rm ext} & \mbox{for} & R_w<1 \\[0.5ex]
\sqrt{\sigma_{\rm ext}} & \mbox{for} & R_w=1 \\[0.5ex]
\sqrt{R_w-1}            & \mbox{for} & R_w>1 \ \
                          \mbox{and}\ \ \sigma_{\rm ext}=0
\end{array}\right.\,.
\label{sigma_y_scaling}
\end{equation}
In the subcritical region one has linear response,
as expected, with the limiting relation 
$\lim_{R_w\to0} \sigma_y = \sigma_{\rm ext}$.
The system is in contrast highly susceptible to 
perturbations at criticality.

The activity resulting from the stationarity
condition (\ref{sigma_y}) is illustrated
in Fig.~\ref{fig_APT_activity}. One observes
the classical phenomenology of an absorbing
phase transition \cite{gros2015complex}, with
the driving input $\sigma_{\rm ext}$ taking
the role of an external field. For the isolated
system, when $\sigma_{\rm ext}\to0$, a transition
from an inactive absorbing phase to an autonomously
active state takes place at $R_w=1$. Finite
fields $\sigma_{\rm ext}$ smooth the transition.

%--------------------------------------
\begin{figure}[t]
\centerline{
\includegraphics[width=0.9\textwidth]{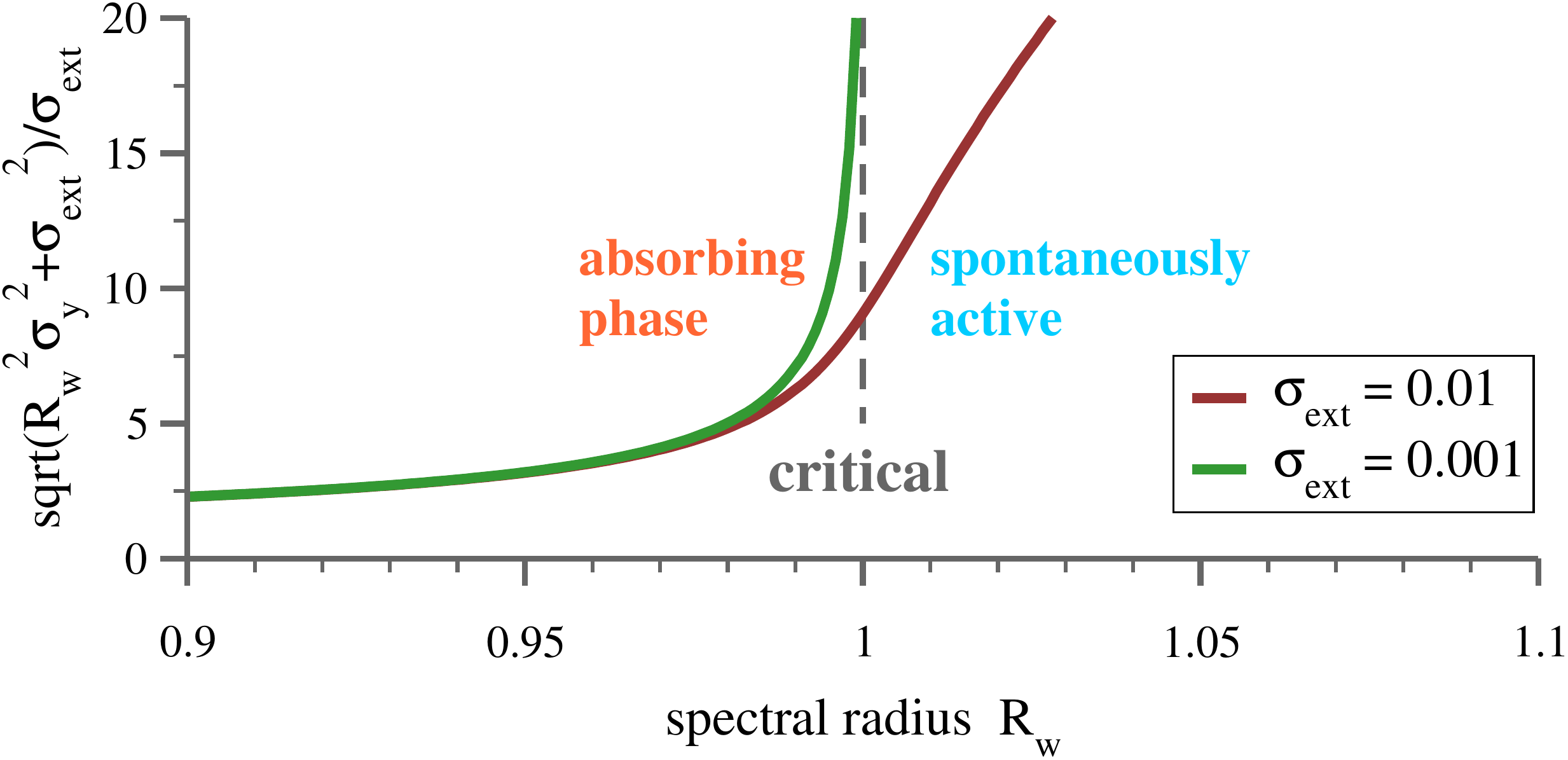}
           }
\caption{{\bf Critically enhanced afferent inputs.}
For the data shown in Fig.~\ref{fig_APT_activity}, the 
enhancement of the neural activity close the critical point.
Given is the ratio of the standard deviation
$\sqrt{R_w^2\sigma_y^2+\sigma_{\rm ext}^2}$ of the
total input neurons receive, with respect to $\sigma_{\rm ext}$.
At criticality, $R_w=1$, the enhancements are
9.07\,/\,28.5, respectively for
$\sigma_{\rm ext}=10^{-2}/10^{-3}$. The enhancement
vanishes, becoming unity, for $R_w\to0$.
}
\label{fig_APT_induced}
\end{figure}
%--------------------------------------

Regarding the brain, of interest is the enhancement
of the induced activity by the recurrent contribution.
A precise quantification of the activity enhancement
is provided by the ratio
\begin{equation}
E_\sigma =
\sqrt{\frac{R_w^2\sigma_y^2+\sigma_{\rm ext}^2}{\sigma_{\rm ext}^2}}\,,
\label{enhancement}
\end{equation}
where $\sqrt{R_w^2\sigma_y^2+\sigma_{\rm ext}^2}$
is the standard deviation of the total input a typical 
neuron receives, and $\sigma_{\rm ext}$ the standard 
deviation of the driving afferent activity. Note that there 
is no enhancement in the absence of recurrent connections, 
viz that $E_\sigma\to1$ when $R_w\to0$. At criticality, when
$R_w=1$, it follows from (\ref{sigma_y_scaling}) that
$E_\sigma$ diverges like $1/\sqrt{\sigma_{\rm ext}}$,
the telltale sign of computation at the edge of chaos 
\cite{langton1990computation,legenstein2007edge}. 
The functional dependence 
is illustrated in Fig.~\ref{fig_APT_induced}. 

Instead of $E_\sigma$ one could equally well consider the difference 
between neural activities with and without external driving, 
which is done by subtracting for $R_w>1$ the autonomous contribution
$\sigma_y\sim \sqrt{R_w-1}$, as listed in (\ref{sigma_y_scaling}).
The resulting susceptibility has a characteristic peak at the 
transition \cite{gros2015complex}.

%---------------------------
\subsection{Spiking vs.\ rate encoding neurons}
%---------------------------

The stationarity condition (\ref{sigma_y}) provides,
modulo residual inter-site correlations, a universal 
and faithful description of driven networks of rate encoding
neurons. The result is that an absorbing phase 
transition is observed, and not any other kind,
ruling out that the brain operates, f.i., at the
edge of synchronization or percolation
\cite{munoz2018colloquium,tinker2014power}. The
operative reason is, intuitively, that adapting
mechanisms need to be able to make excursions to
both sides of the transition line. Both the sub- and
the over-critical state must hence be biologically 
viable. One may ask whether this insight is particular 
to networks of rate encoding neurons. This seems
not to be the case. 

It has been shown recently, that generalized, non-conserving 
sandpile models undergo absorbing phase transitions \cite{gobel2020absorbing}.
A close correspondence between toppling and spiking events is 
present in these models, which strongly suggest that
transitions from inactive to autonomously ongoing neural
dynamics follows the phenomenology of absorbing phase 
transitions also for spiking networks.

%---------------------------
\subsection{Absorbing phase transitions vs.\ branching processes}
%---------------------------

Transitions from inactive to active neural dynamics are
separated by an absorbing phase transition. This
holds, as argued above, generically, in particular
both for networks of spiking and rate-encoding neurons.
Alternatively, it has been argued, that close-to critical
brain activity shares features with branching processes
\cite{haldeman2005critical}. This would be somewhat surprising, 
if true, given that the activity of a super-critical branching 
process strictly diverges. For the sake of terminology
is may hence be worth pointing out that what has been denoted in 
\cite{haldeman2005critical} a branching model, corresponds 
in reality to a discretized stochastic neural network which 
undergoes, as expected, an absorbing phase transition. 
There is nevertheless a close analogy between subcritical
branching processes and the dynamics leading to an absorbing
state \cite{dickman1998self}. Both processes die out 
exponentially, with the respective decay times diverging 
when approaching the critical point. Regarding
the associated universality classes, it has been established 
that absorbing phase transitions belong in many cases to 
the universality class of directed percolation 
\cite{gobel2020absorbing,lubeck2005scaling}. For the brain,
the large degree of inter-connectivity corresponds to 
high-dimensional lattice models, which follow mean-field 
scaling.

%===========================
\section{Self-sustained and resting state brain activity}
%===========================

Our brains use a non-negligible amount of
energy, about 20\% of what the body consumes
\cite{clarke1999circulation}, for `doing nothing', 
to say, namely for the upkeep of its resting 
state. Engaging in tasks needs comparatively little 
additional energy, of the order of 5\% of the resting 
state consumption \cite{raichle2006brain}. This would
a trivial observation if the energy demand of the brain
would be dominated by basic metabolism, just to keep 
our wetware alive, which is however not the case. The 
larger part of the energy consumed by the brain involves 
information processing \cite{raichle2009paradigm}.
Energy balance considerations suggest then that human 
brains are concerned  mostly with themselves, a view that 
is reinforced by the observation that only a compartively 
small number of synapses is devoted to stimuli 
proccessing \cite{raichle2015restless}.

%---------------------------
\subsection{Is the brain driven or modulated?}
%---------------------------

Various views regarding the role of the resting state activity
are viable \cite{northoff2010brain,deco2011emerging}. 
A first one would be to assume that the primary task 
of the ongoing autonomous neural dynamics is to 
prepare the brain for the processing of external 
stimuli. This view is often implicitly
assumed when positing that the brain operates
lose to criticality \cite{beggs2012being}, given that 
working regimes in the vicinity to a phase transition 
come with favorable statistical properties
for the processing of external inputs
\cite{bertschinger2004real}.

As a second possibility one may take the existence 
of a substantial resting state activity as an 
indication that the brain is not a driven systems, 
in terms of stimulus-response functionalities,
but a modulated organ \cite{fiser2004small,gros2009cognitive}. 
The ongoing and to most parts self-sustained activity 
involves in this view non-trivial information processing, 
which is in general redirected, viz modulated by incoming 
sensory stimuli 
\cite{rabinovich2001dynamical,kenet2003spontaneously},
but not driven. Only stimuli of exceptionally large 
intensities would force subsequent brain areas
to specific responses. Sensory inputs interact in 
this view with the ongoing transitions between 
internal cognitive states 
\cite{gros2007neural,silvanto2008state},
which means that perception induces 
transitions among neuronal attractor states 
\cite{braun2010attractors}.

A candidate framework for the modulated interaction
between brain and outside world is semantic learning, 
for which pre-existing internal states acquire 
semantic meaning while interacting with the environment
\cite{gros2010semantic}. This process is ongoing,
viz while the brain is autonomously sampling available
internal states. Learning occurs when sensory stimuli
redirect the sequence of neuronal states. Experimentally, 
there are indeed indications that the brain acts as a 
nonlinear dynamical system, with trajectories that are 
tighten when performing a task \cite{he2013spontaneous}.

The distinction between modulated and driven systems
is, strictly speaking, not rigorous. In both cases 
sensory stimuli leave in the end an imprint on the brain. 
It is however a question of starting points. In physics, 
to given an example, one can develop the theory of a 
non-ideal gas starting either from zero temperature, 
or from the high-temperature limit. Depending which 
regime one wants to describe, either can be the better
starting point. The same hold in our view for
the role of the resting state activity. 
We posit here that the appropriate starting 
point for the brain is the hypothesis that the 
autonomous activity involves non-trivial internal
computations which are essential to the functioning
of the brain, and not just an side effect.

As an interesting twist one could speculate which
ingredients are necessary for an hypothetical,
silicon-based artificial brain. Assuming that
such an AI would be based on biological correspondence
principles, the question would be whether an
operational regime close to criticality is
just performance enhancing, or a conditio sine
qua non, with the first option corresponding to the
viewpoint that the brain is in essence a
reflexive organ.

%---------------------------
\subsection{Stationary resting state activity}
%---------------------------

There are good reason to assume, as pointed out
above, that the resting state activity of the
brain has intrinsic functionality
\cite{reineberg2015resting}, viz that it is 
doing more than to help the brain to process 
incoming stimuli somewhat more efficiently. See,
e.g., \cite{papo2013should} for a review.
Following the stationarity principle, this implies
that the overall amplitude of the autonomous activity 
needs to be regulated. This can be achieved for an
arbitrary spectral radius $0<R_w$ when tuning up the 
afferent synaptic weights correspondingly, namely
by selecting an appropriate $\sigma_{\rm ext}$. 
Compare Fig.~\ref{fig_APT_activity}. The desired 
internal activity level could be generated in this 
way, at least as a matter of principle. There are 
however two caveats. 

The first point regards the balance between
invoked and internally generated activity, 
given that an $R_w$ substantially lower than unity 
is characteristic for driven systems. There would
be no room for a non-trivial resting-state dynamics.
The second caveat is that the intensity of the input, 
quantified in (\ref{sigma_y}) by the average standard 
deviation $\sigma_{\rm ext}$, is not constant. Consider
as an example the task dependency of the 
activity levels of specific brain areas, which
can be in part substantial 
\cite{tagliazucchi2011spontaneous}. These are 
typically local enhancements with only limited 
impact on the overall energy consumption
of the brain in its entirety \cite{raichle2006brain}. 
Brain areas project however further on, which
implies that the respective downstream areas have 
to deal with varying levels of input intensities,
which is arguably a generic feature of neural circuits.
In the end, the only venue to attain neural activity
that is generated internally, at least in good part,
is a spectral radius close to unity. This view
is consistent with an analysis of in vivo spike 
avalanches that indicates that the brain is
slightly subcritical \cite{priesemann2014spike}.

%---------------------------
\subsection{Regulated or self-organized criticality?}
%---------------------------

It has always been controversial, what the
term `self organization' exactly means
\cite{sornette2006critical}. Roughly speaking,
it implies that novel, somewhat non-trivial
properties emerge from basic rules
\cite{gros2015complex}. In particular
one would like to see that the emerging properties 
do not result from a straightforward 
``get out what you put in'' mechanism.
These consideration definitively hold
for the original concept of self-organized
criticality by Bak {\it et al.} \cite{bak1987self}, 
for which power-law scaling results from two 
ingredients: infinite time scale separation and 
energy conservation \cite{markovic2014power}.
The situation is less clear when it comes to
the explicit regulation of a system close to
the point of a phase transition. The resulting
system will be critical, but clearly not in
a `self-organized' critical state
\cite{bienenstock1998regulated}. An example
for a basic rule allowing a neural network to
adapt towards criticality is flow control,
as defined by (\ref{flow_stationary}). A state
close to criticality may be achieved also 
indirectly by tuning in first place the 
EI (exciation/inhibition) balance 
\cite{trapp2018ei,ma2019cortical}.

%===========================
\section{Routes towards criticality}
%===========================

When it comes to evolve neural networks towards 
criticality, all proposed methods regulate in 
the end the flow of activity through the network.
Whether the flow contracts or expands is determined
by the spectral radius (\ref{R_w}) of the propagator
in question, which is typically composed of the synaptic 
weight matrix modulo neuronal rescaling factors, such
as the gain. The various conceivable routes for achieving
stationarity, either directly or indirectly, can 
be classified along several criteria:
\begin{itemize}
\item {\bf direct vs.\ indirect}
\item {\bf online vs.\ offline}
\item {\bf adapting synaptic weights vs.\ other parameters}
\end{itemize}
Direct methods control the flow explicitly, indirect 
approaches do not. Online routes towards criticality
can be implemented while the system is operative, 
ideally while processing afferent signals. Offline 
algorithms consider on the other side isolated systems. 
The prime target for adaption is often the synaptic 
weight matrix, a procedure which could interfere however 
with internal Hebbian-type learning. Other parameters 
to be adapted are either global, e.g.\ in the context of
neuromodulation, or local. Examples for the latter case
are individual neuronal gains and thresholds. Of importance
is also how long it takes to close in towards criticality,
at least in order of magnitude. Note in this regard that 
the time scales for compensatory mechanisms in the brain 
are often faster in models, than experimentally reported 
\cite{zenke2017hebbian}. On the backdrop of these general 
considerations, we discuss now some representative examples, 
albeit without aiming for completeness. For an overview
regarding the role of distinct types of synaptic plasticities
see \cite{zeraati2020self}.

A direct and potentially online method is anti-Hebbian 
tuning \cite{magnasco2009self}, which acts expressively
on the synaptic weight matrix. Time is continuous, which
implies that the real part of the eigenvalues of the
synaptic weight matrix matter, as laid out in 
Sect.~\ref{sect_activity_flows}. Interestingly, not just 
the real part of the largest, but of all eigenvalues
are forced to adapt to zero, which is not the case
for standard echo-state frameworks. For anti-Hebbian
tuning non-local information is necessary, namely the 
cross correlation of arbitrary pairs of neurons.

An example for an indirect algorithm is a proposal
by Bornholdt and Rohlf \cite{bornholdt2000topological}, 
which makes use of the properties of the attracting
states emerging in fully isolated systems. The update
rules of the connection matrix are dependent in this
scheme on the properties of the observed attractors 
\cite{meisel2009adaptive}. 
The method is hence offline, with an diverging timescale 
for the adaption process, which may be in conflict
with the experimentally observed lack of neuronal
time-scale separation
\cite{priesemann2014spike,das2019critical}.

A recently investigated abstract mechanism makes use of
the fact that order-parameter fluctuations, in essence
the susceptibility, peak at the locus of a second order 
phase transition \cite{chialvo2020controlling}. Adapted 
is a global parameter, the temperature. For the estimates 
of the order-parameter fluctuations non-local information
is necessary. A generalization to systems with external
fields, viz inputs, should be possible.

Adaption mechanisms for fully operative systems, viz 
for networks continuously processing incoming signals,
are a rare species. An example is flow control, which 
regulates the flow of activity through individual neurons, 
using an adaption scheme that is based on locally 
available information. Regulating intrinsic neuronal
parameter, typically the gain \cite{schubert2020local},
an online method is obtained. Here the spectral radius 
$R_w$ enters expressively the stationarity condition 
for the neuronal activity, as given by (\ref{flow_stationary}).

%---------------------------
%\subsection{Critical neurons?}
%---------------------------

%===========================
\section{Discussion}
%===========================

It is tempting to assume that enhanced
information processing \cite{kinouchi2006optimal,shew2009neuronal}
is the prime evolutionary driver for cortical
networks to evolve close to a critical point
\cite{chialvo2010emergent}.
We have pointed here to an alternative, namely
the need to maintain a non-trivial level of
autonomously generated neural activity. To this 
regard several adaption mechanisms encoding 
the evolution of an operative system towards
the onset of a second order phase transition
have been proposed, by monitoring either the
intrinsic activity, as for flow control 
\cite{schubert2020local}, or higher-order
correlation functions \cite{chialvo2020controlling}.
A common precondition is the existence of a 
non-vanishing ongoing activity that is not 
induced in its entirety by external inputs.
The rational is straightforward. Internally
generated activity is directly impacted by
a nearby critical point, evoked activity in
contrast only indirectly -- which loops
back to the starting point, the assumption
that the resting state activity has key 
functionalities. To which extend this
assumption holds needs in our view further 
efforts regarding the study of the interplay 
between the autonomous brain dynamics and 
cognitive information processing. Present
data seem to support the notion that ongoing 
neural activity does not just encode statistically
structured noise, but multidimensional behavioral 
information \cite{stringer2019spontaneous}.

%===========================
\section*{Acknowledgments}
%===========================

The author thanks Dante Chialvo and
Fabian Schubert for extensive stimulating discussions.

%===========================
\section*{Data availability statement}
%===========================

Any data that support the findings of this 
study are included within the article.

%===========================
\section*{References}
%===========================

%%%%%%%%%%%%%%%%%%%%%%%%%%%%
\bibliographystyle{unsrt}
%\bibliographystyle{pnas-new.bst}
%\bibliography{devilsAdvocate.bib}

%%%%%%%%%%%%%%%%%%%%%%%%%%%%

\end{document}